\documentclass[fleqn,usenatbib]{mnras}

\usepackage{newtxtext,newtxmath}

\usepackage[T1]{fontenc}

\DeclareRobustCommand{\VAN}[3]{#2}
\let\VANthebibliography\thebibliography
\def\thebibliography{\DeclareRobustCommand{\VAN}[3]{##3}\VANthebibliography}

\usepackage{graphicx}	

\newcommand{\about} {$\sim$}

\title[Jaffa cakes are cakes]{Using Artificial Intelligence to Shed Light on the Star of Biscuits:\\The Jaffa Cake}

\author[H. F. Stevance]{
H. F. Stevance,$^{1}$\thanks{E-mail: hfstevance@gmail.com}
\\
$^{1}$The Department of Physics, The University of Auckland, Private Bag 92019, Auckland, New Zealand\\
}

\date{Accepted XXX. Received YYY; in original form ZZZ}

\pubyear{2021}

\begin{document}
\label{firstpage}
\pagerange{\pageref{firstpage}--\pageref{lastpage}}
\maketitle

\begin{abstract}
Before Brexit, one of the greatest causes of arguments amongst British families was the question of the nature of Jaffa Cakes. Some argue that their size and host environment (the biscuit aisle) should make them a biscuit in their own right. Others consider that their physical properties (e.g. they harden rather than soften on becoming stale) suggest that they are in fact cake. In order to finally put this debate to rest, we re-purpose technologies used to classify transient events. We train two classifiers (a Random Forest and a Support Vector Machine) on \about 100 recipes of traditional cakes and biscuits. Our classifiers have 95 percent and 91 percent accuracy respectively. Finally we feed two Jaffa Cake recipes to the algorithms and find that Jaffa Cakes are, without a doubt, cakes. 
Finally, we suggest a new theory as to why some believe Jaffa Cakes are biscuits.
\end{abstract}




\section{Introduction}
\label{sec:intro}
\begin{figure}
\centering
    \includegraphics[width=6cm]{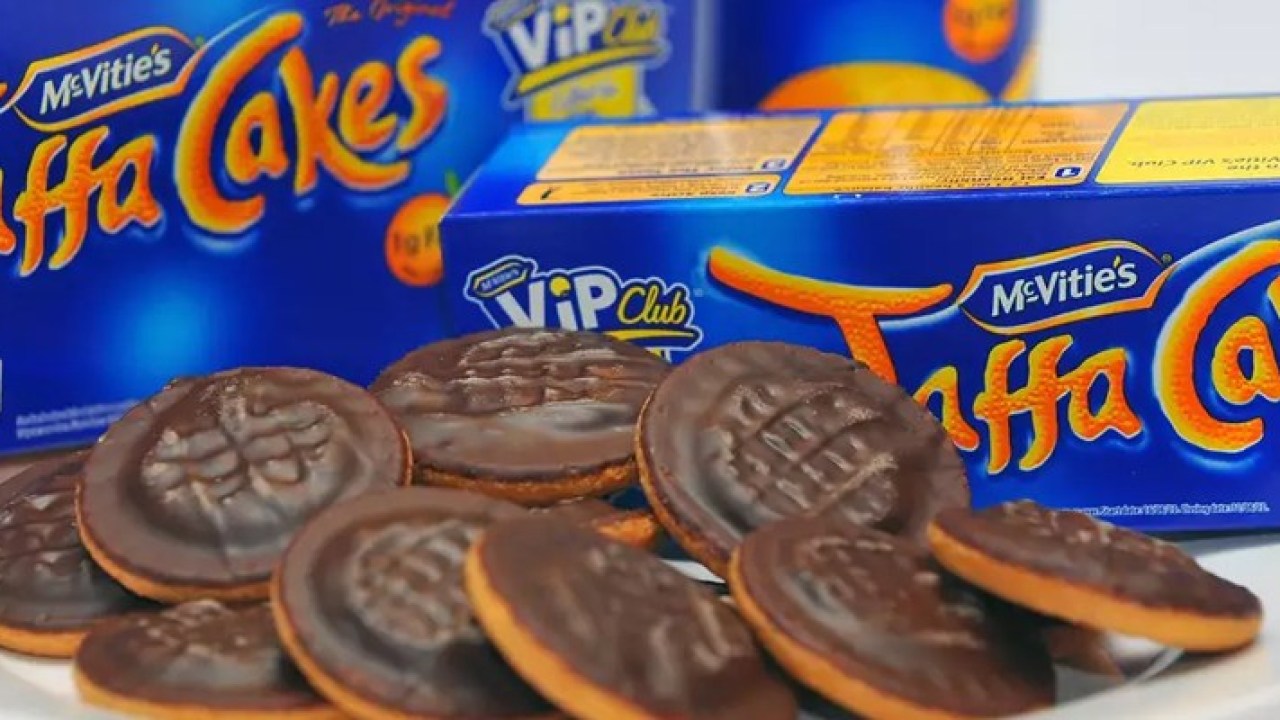}
    \caption{\label{pic:jaffa}Jaffa Cakes}
\end{figure}

\begin{figure}
\centering
    \includegraphics[width=2.5cm]{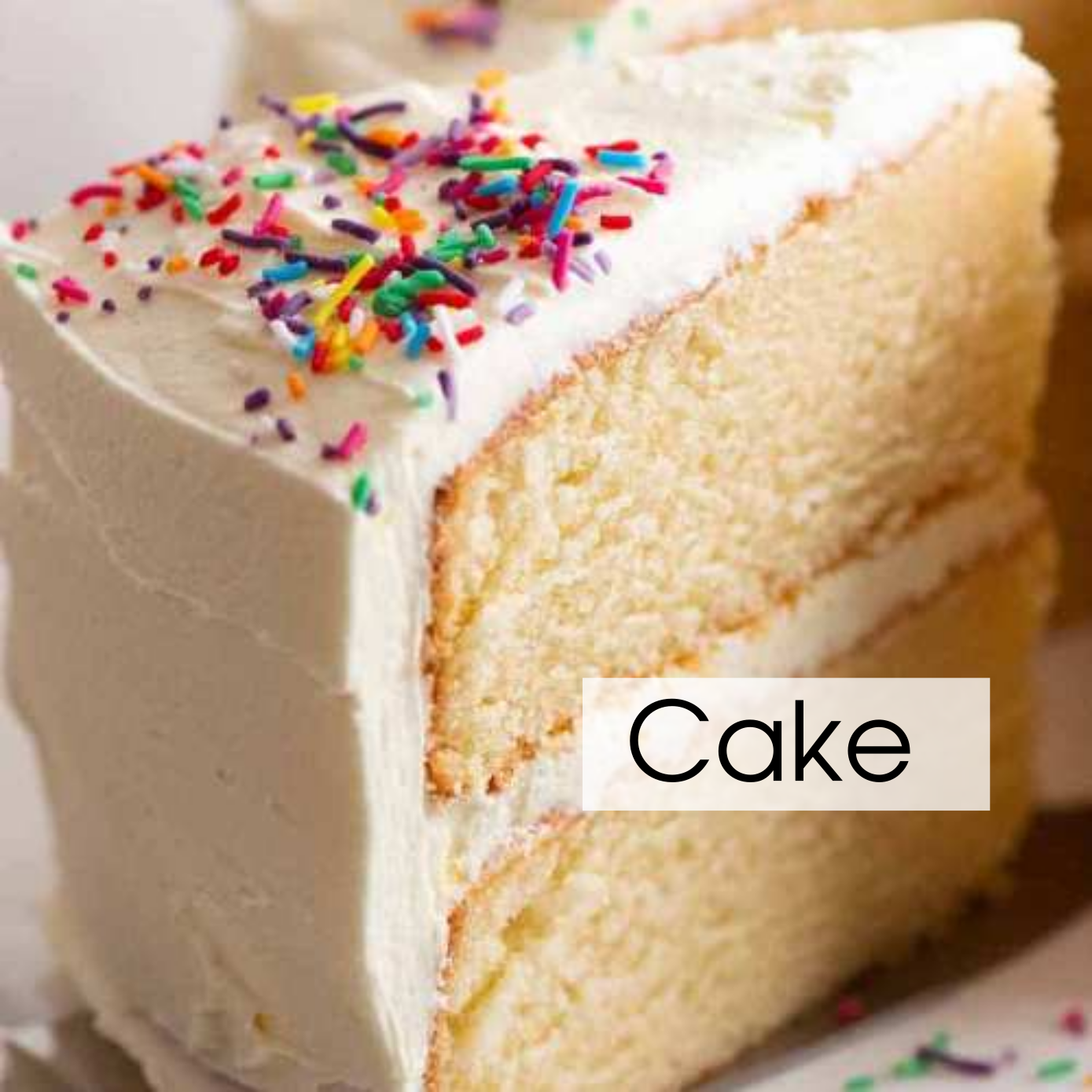}
    \includegraphics[width=2.5cm]{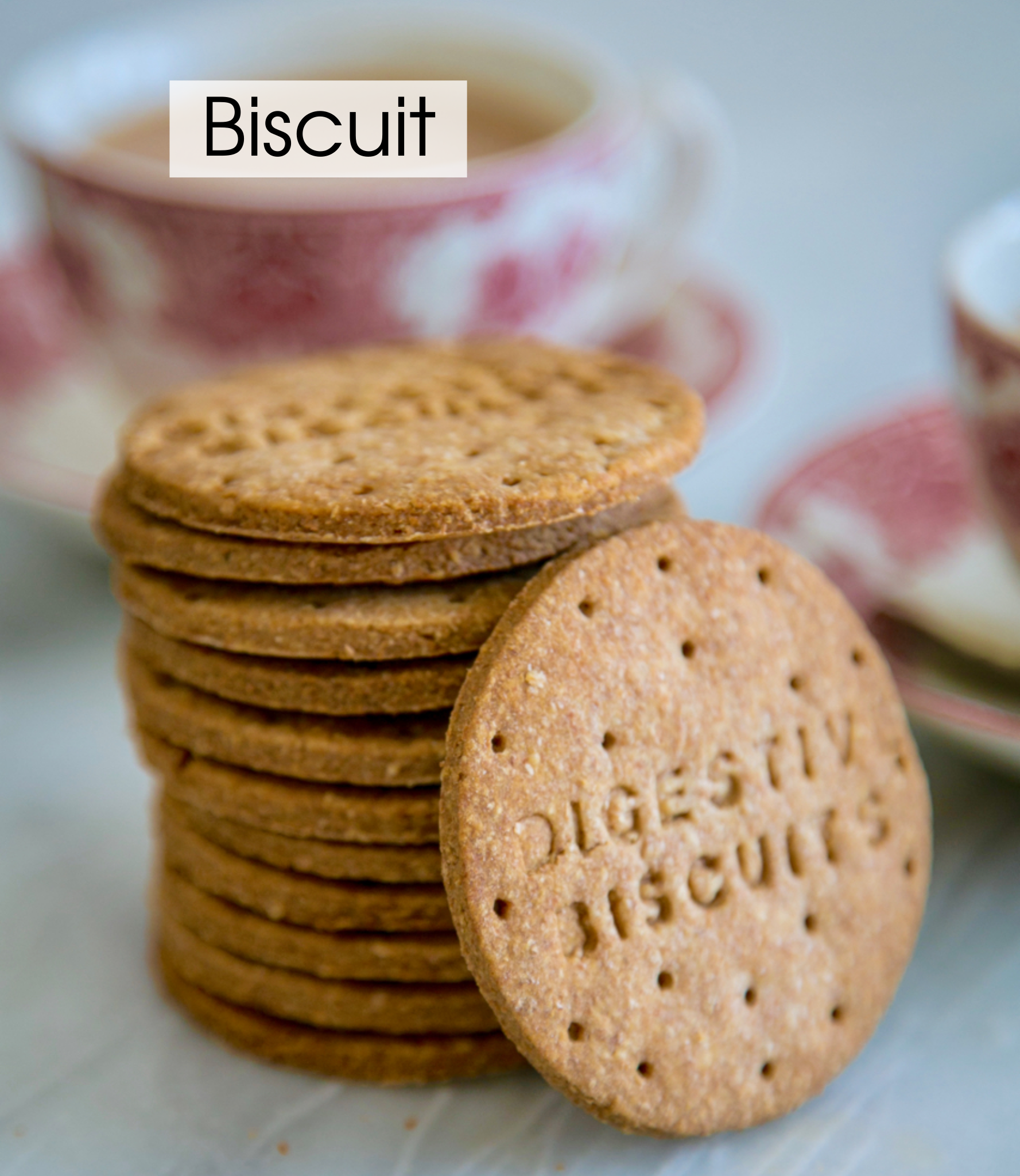}
    \includegraphics[width=2.5cm]{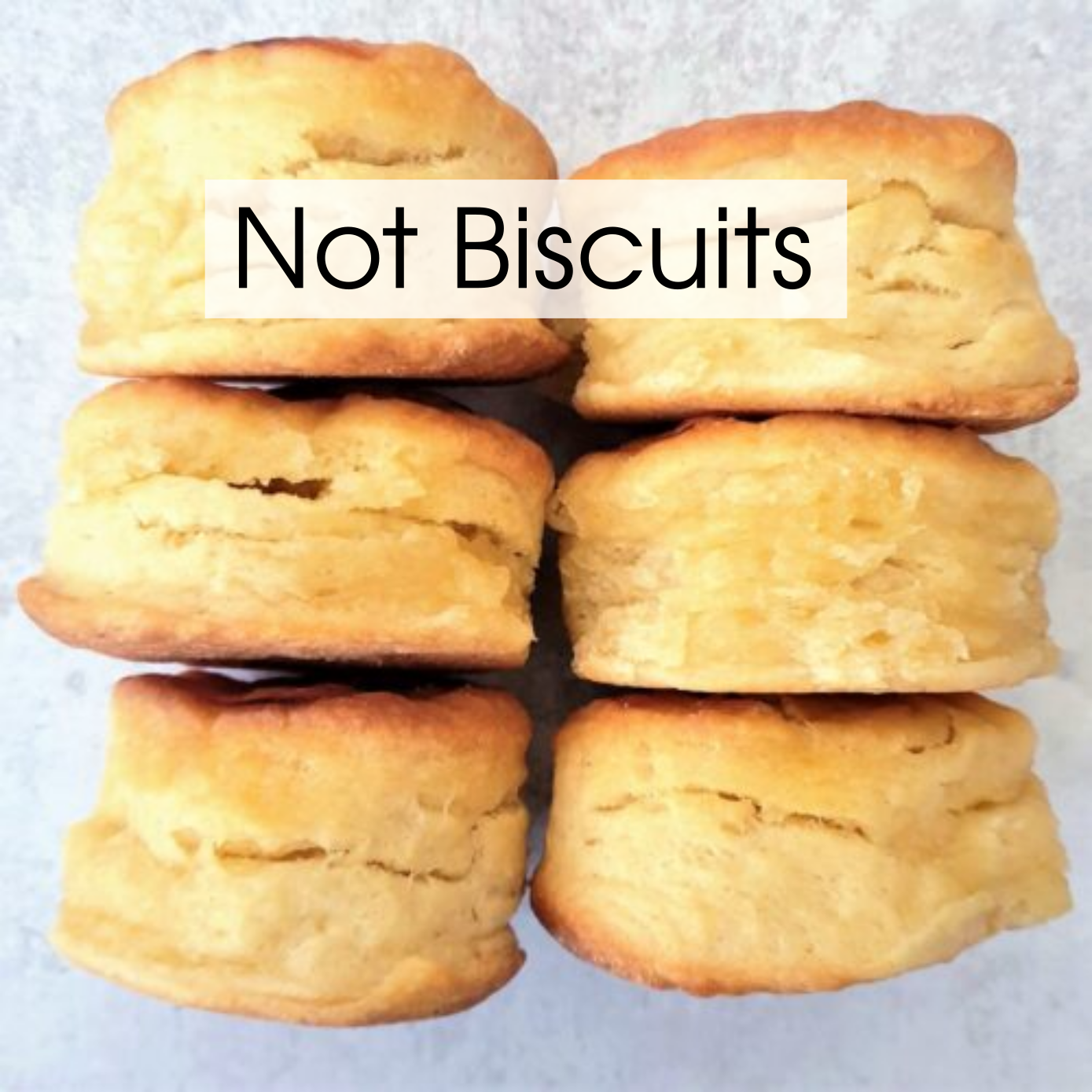}
    \caption{\label{pic:cakes} Visual examples of the cake and biscuit class. It is important to note that scones are not biscuits, get it right USA. }
\end{figure}

``Jaffa Cakes" (Figure \ref{pic:jaffa}) are a popular treat across the United-Kingdom, often eaten alongside tea and served amongst Digestives, Hobnobs, Bourbons and Jammie Dodgers [1, 2].
They are made of three layers: a Genoise sponge topped with a layer of jelly, all covered in chocolate [1]. 
Since their creation nearly 100 years ago [3], their existence has sparked heated debate amongst British families: Are ``Jaffa Cakes" biscuits or cakes? 
Before providing further details of the debate surrounding the orange-flavoured delicacy, it is important to specify our taxonomy: In Figure \ref{pic:cakes} we show examples of the Biscuit and Cake class -- we also highlight an example of something that is definitely not a biscuit, but most assuredly a scone [4]. 

Now that proper terminology has been established, it is worth reminding the reader of the current state of the field:
On the one hand, Jaffa Cakes are sold in the biscuit aisle, measure only 2.125 inches in diameter, and are eaten alongside (or in place of) biscuits [1--3]. 
This evidence has brought some people to argue that Jaffa Cakes should be considered a biscuit in their own right. 
However, a stale Jaffa Cake tends to harden, similarly to a left-over piece of cake; on the contrary biscuits tend to soften on going stale.
These observations have a solid history, and repeated experiments have found very consistent results [5, 6]. 
This would suggest that Jaffa cakes are \textit{not} biscuits, but small cakes.

The question of the nature of the Jaffa cakes is therefore not a trivial one but it has been argued in court in 1991 by McVities that they are indeed cake and not biscuits. 
It has been reported though that this work was not scientifically motivated and simply a ploy to avoid paying VAT. This evidence is therefore often rejected by those who argue that Jaffa Cakes are biscuits [7]. 

This debate is therefore very much alive. The aim of this paper is to finally put it to rest by using Artificial Intelligence.
To this end I gathered \about 100 recipes and trained a binary classifier to recognise cakes and biscuits, after-which I fed a Jaffa Cake recipe into the trained algorithms for classification. 

In Section \ref{sec:data} I describe the data set and how it was gathered. In Section \ref{sec:results} I provide details on the training and usage of the binary classifiers. Finally in Section \ref{sec:disc} we discuss and conclude.

\section{Data set}
\label{sec:data}

The data set was created by gathering 51 recipes of biscuits, 41 recipes of cakes and 2 recipes of Jaffa Cakes from the internet. 
The data used to train and test the algorithm only contained the labeled cake and biscuit recipes, leaving out the Jaffa Cakes which have yet to be classified. 
It is important to note that one of the Jaffa Cake recipes is from Mary Berry [8] -- the other one is very similar. Therefore, they are beyond questioning and we will not be taking constructive criticism on the small sample size.

A wide range of cakes and biscuit recipes were included in the sample and each ingredient is considered a feature. 
Additionally, we created the `WET', `FAT' and `DRY' features by summing the weights of the wet ingredients (e.g. egg, milk), sources of fat (e.g. butter, oilf), and dry ingredients (e.g. flour, chocolate powder, oats). 
Each  raw feature had to fit into one of the three engineered features such that the total weight of the recipe was equal to the combined weight of `WET', `FAT' and `DRY'. 

Finally, we normalised each recipe by total weight and then created one final feature: the Wet-To-dry Fraction, or WTF index for short. The WTF index quantifies the wetness of the recipe - from experience this seems to be one of the distinguishing features between biscuit doughs and cake batters. A similar criterion was used by the Irish Revenue Commissioners to determine what VAT to apply to Jaffa Cakes [9].

\section{Training The Algorithms and classification}
\label{sec:results}

\subsection{Random Forest}
We first trained a Random forest classifier with 50 decision tree; using k-fold cross validation we determine an accuracy of 95 percent. 
After training the algorithm we can look at the relevance of each feature -- the relevance score returned by {\tt sklearn} is recorded in Table \ref{tab:relevance} and we discuss the most important features below.

\begin{table}
    \caption{\label{tab:relevance} Feature relevance from Random Forests}
    \centering
        \begin{tabular}{lr}
        \hline
        Feature &  Importance Score \\
        \hline
        Eggs              &  0.239093 \\
        Self-raising flour           &  0.158565 \\
        WTF                &  0.143973 \\
        FAT                     &  0.096131 \\
        DRY                     &  0.087810 \\
        Milk or buttermilk (ml) &  0.081340 \\
        Caster sugar            &  0.055094 \\
        Plain flour             &  0.052134 \\
        Oil                     &  0.016414 \\
        Brown sugar             &  0.015169 \\
        WET                     &  0.011714 \\
        Oats                    &  0.010530 \\
        Baking soda (tsp)       &  0.009678 \\
        Golden syrup            &  0.006932 \\
        Butter                  &  0.006569 \\
        Baking powder (tsp)     &  0.004130 \\
        Ground almond           &  0.003853 \\
        Cocoa powder            &  0.000870 \\
        \hline
        \end{tabular}
\end{table}

\subsubsection{Eggi-ness}
The egg quantity was by far the most informative feature (0.239). Naively this may be interpreted as being a consequence of the wetness of the egg. Indeed, moisture has been used in the past to classify the Jaffa Cake and the WTF index scores high in relevance (0.144 -- see Table \ref{tab:relevance}). 
However, the  WET feature (the fraction of wet ingredients in the recipe by weight) alone scores quite low (0.012) and therefore we must conclude that the wetness of the egg is not what causes it to be so informative for our classifier. 

The relevance of the egg is more likely to stem from the fact that it is a binding agent necessary to the structural integrity of cakes, and often omitted in biscuits. Shortbread for example requires no egg [10] - its structural integrity relies on a thick flour-rich dough. More liquid cake batters will need some binding agent that hardens on baking [11]. In non-vegan recipes this role is most often filled by the egg [11, 12]. 

\subsubsection{Self-raising flour}
We disregard this particular feature because biscuit recipes, although they often do contain raising agents, tend to add the raising agent separately rather than use self-raising flour which is most often used in cakes [13]. 
Conversely, some cakes need no raising agent and rely on cooking method to obtain a cake-y consistency (e.g. by beating the egg whites stiff) [14]. 

\subsubsection{WTF}
\begin{figure}
\centering
    \includegraphics[width=7cm]{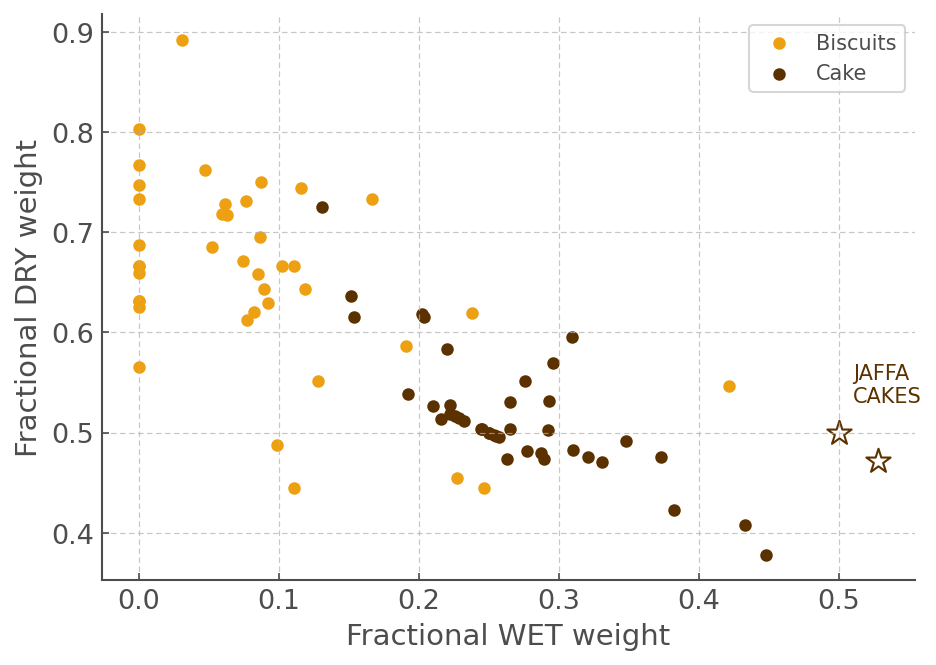}
    \caption{\label{pic:wet-dry}Our labeled sample in the DRY-WET parameter space.}
\end{figure}

The WTF index indicates how moist (or wet) the mixture will be and is another very informative feature as to whether a recipe will produce a cake or a biscuit. 
We can visualise this by plotting our labeled data in a DRY-WET parameter space, see Figure \ref{pic:wet-dry}.

We can see that most biscuits are very dry (WET<0.1 and DRY>0.6), whereas most biscuits are very wet (WET>0.25 and DRY<0.6).
The region 0.1<WET<0.25 contains both cake and biscuits but mostly cakes. This can be seen as a transition region where ingredients alone is not sufficient to determine the nature of the baked good. 
Method and baking time will also have an influence and have been ignored here. 

Finally the biscuit with WET>0.4 is considered an outlier. The most likely explanation is that I messed up when I typed the quantities into my excel sheet. 

\subsubsection{Classifying the Jaffa Cakes}

When classifying the Jaffa Cake recipes, our Random Forest algorithm predicted them to have the label "Cake" with 100 percent certainty. Given that our model is 95 percent accurate that means we can be 95 percent sure that Jaffa Cakes are cakes \footnote{stats is hard}. 

\begin{figure}
\centering
    \includegraphics[width=7cm]{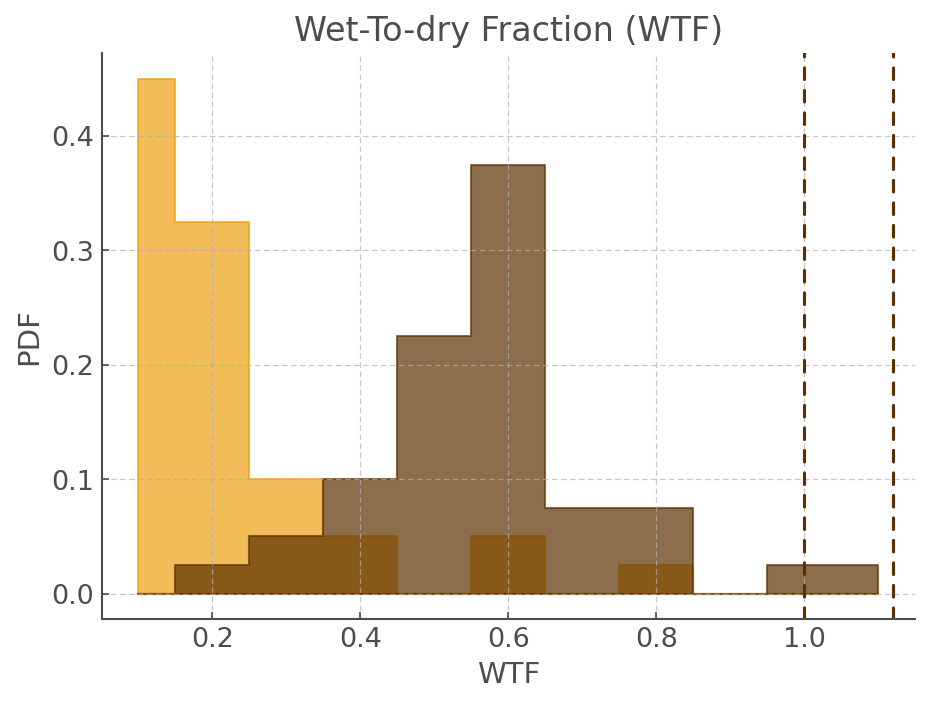}
    \includegraphics[width=7cm]{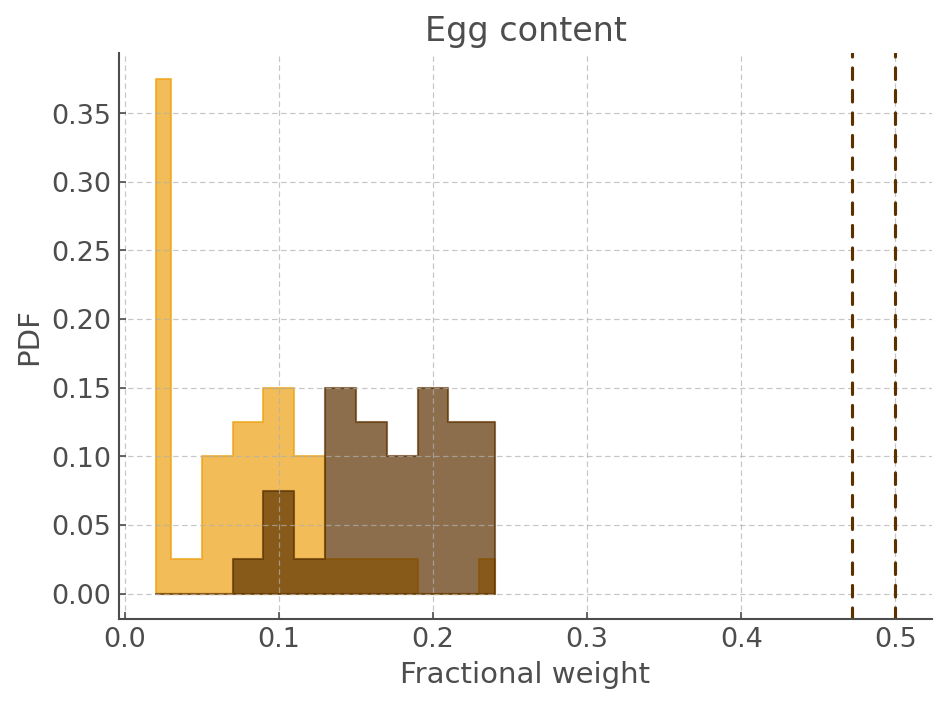}
    \caption{\label{pic:wtf} Distribution of the WTF index and egg content in our sample. The Jaffa Cake values are shown by the vertical dashed lines.}
\end{figure}

We can verify this result by comparing the features of our Jaffa Cake recipes to that of our labeled data set. The WTF index of the Jaffa Cake recipes exceeds the 0.95 percentile and the egginess is over double the maximum observed in our Cake sample (see Figure \ref{pic:wtf}. This is already overwhelming evidence that Jaffa Cakes are cakes, but Random Forests are boring [15] and I want to show off the things I learnt on the internet so in the next section we will use Support-Vector Machines.

\subsection{Support-Vector Machine}
\begin{figure}
    \includegraphics[width=\columnwidth]{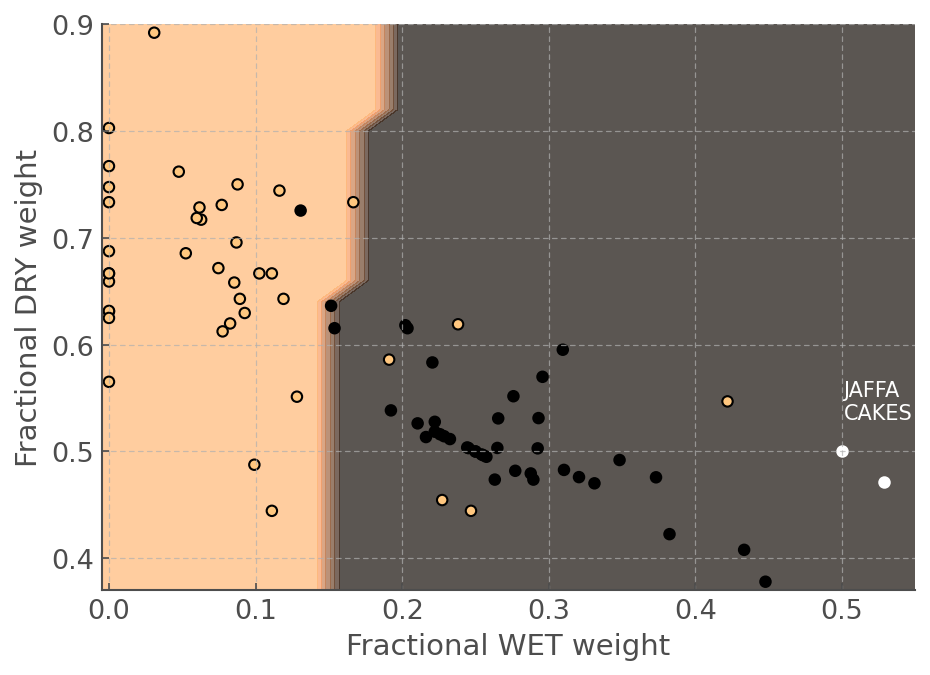}
    \caption{\label{pic:svm}SVM classification in DRY-WET space. Brown is the cake class, orange is the biscuit class.}
\end{figure}

Using a 2nd degree polynomial kernel we can separate the cakes and biscuits in DRY-WET space with 91 percent accuracy -- see Figure \ref{pic:svm}. We find that less than 10 (2) percent of the biscuits (cakes) are miss-classified. Once again the Jaffa Cake recipes are classified as Cake. 

There may or may not be caveats to this classifier -- take a look at Appendix \ref{sec:A1.2} at your own peril.

\section{Discussion and Conclusion}
\label{sec:disc}
In this paper we used 51 recipes of biscuits and 41 recipes of cakes to train two binary classifiers: A Random Forest and a Support Vector Machine. 
The two trained algorithms have accuracy 95 and 91 percent respectively, and they both found Jaffa Cakes to be unambiguously cakes. 

Now the author can anticipate that the people claiming that Jaffa Cakes are biscuits will say that this analysis is incomplete because it does not take into account the host environment (biscuit aisle) or size of the Jaffa Cake.
However it is important to point out that size does not the biscuit make. Indeed cupcakes are cakes despite their small stature. Additionally, many a Mr. Kipling cake product -- French Fancies, Angel Slices and Lemon Slices just to name a few -- are also miniature, and they are also sold in the biscuit aisle and enjoyed with tea.
But it has never been argued that a Lemon Slice could be a biscuit. 

Consequently, we consider the host environment to have little baring on the nature of the Jaffa Cake.
It is also worth wondering why the mini-cakes mentioned above are not the object of controversy whilst the Jaffa Cake is. 
One of the main differences between a Jaffa Cake and a traditional cake product, whether it be small and whole (cupcake) or large and sliced (normal slice of cake), is that the Jaffa Cake can be held and dipped without leaving much residue on the fingers. 
Most cakes will be sticky.
As a result, the \textit{eating experience} of the Jaffa Cake will be more similar to that of a traditional biscuit. 

Although this new theory is interesting in understanding the humans eating the Jaffa Cake, it does not in any way detract from its true nature: Jaffa Cakes are cakes.

\section*{Acknowledgements}
I would like to thank my fear of failure for fuelling the streak of procrastination that led to this work.
To all my Sheffield mates : 1) I was right 2) you were wrong 3) you're dead to me. 



\bibliographystyle{mnras}
Google is free.

\vspace{10cm}
\appendix
\section{Beyond the joke}
\label{sec:A1.2}
\begin{figure}
    \includegraphics[width=\columnwidth]{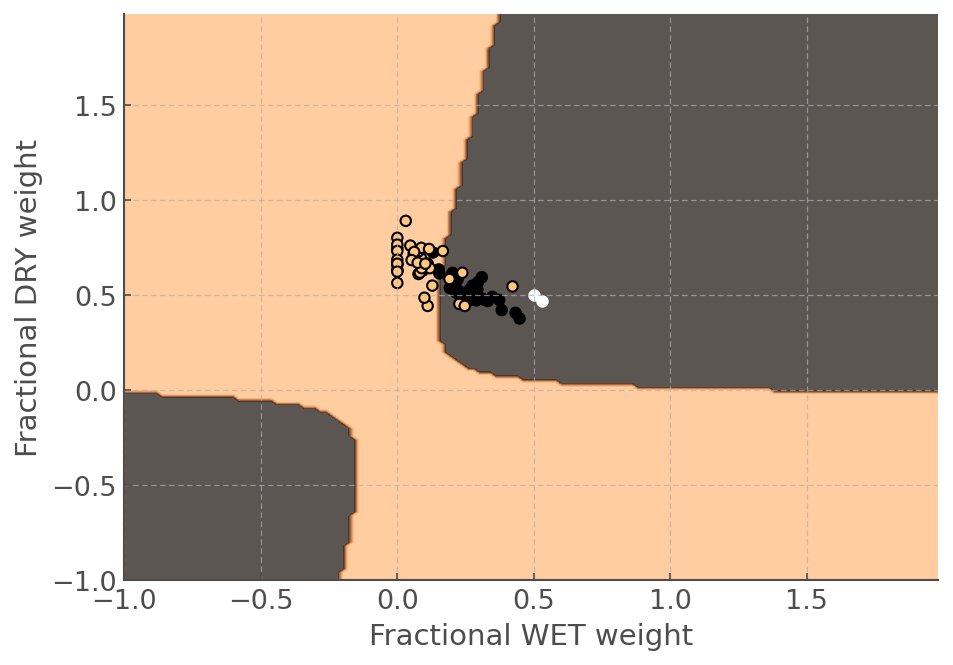}
    \caption{\label{pic:A1}go home A.I. you're drunk}
\end{figure}

The SVM I made is a great example of why you should NEVER use your trained algorithm on data outside the range covered by your training sample. Take a look at Figure \ref{pic:A1}






\bsp	
\label{lastpage}
\end{document}